# Ultrathin plasmonic Ag nanogratings for rapid and highly-sensitive detection

Beibei Zeng[a)], Yongkang Gao, Filbert J. Bartoli[b)]
*Electrical and Computer Engineering Department, Lehigh University, Bethlehem, PA 18015, USA.*
*E-mail:* a) bez210@lehigh.edu; b) fjb205@lehigh.edu

**Abstract** We developed a nanoplasmonic sensor platform employing the extraordinary optical properties of one-dimensional nanogratings patterned on 30*nm*-thick ultrathin Ag films. Excitation of Fano resonances in the ultrathin Ag nanogratings results in transmission spectra with high amplitude, large contrast, and narrow bandwidth, making them well-suited for rapid and highly-sensitive sensing applications. The ultrathin nanoplasmonic sensor chip was integrated with a polydimethylsiloxane (PDMS) microfluidic channel, and the measured refractive index resolution was found to be $1.46 \times 10^{-6}$ refractive index units (RIU) with a high temporal resolution of 1 *sec*. This compares favorably with commercial prism-based surface plasmon resonance sensors, but is achieved using a more convenient collinear transmission geometry and a significantly smaller sensor footprint of $50 \times 50 \mu m^2$. In addition, an order-of-magnitude improvement in the temporal and spatial resolutions was achieved relative to state-of-the-art nanoplasmonic sensors, for comparable detection resolutions.

**Introduction**

Optical biosensors employing the surface plasmon resonance (SPR) in flat thin metal films have unquestionable advantages for *real-time* and *label-free* detection of biomolecular interactions, with detection resolution in the range between $10^{-6}$ and $10^{-7}$ RIU and a temporal resolution of 1 *sec* [1-2]. Over the past two decades, SPR sensors have become the accepted standard for the rapid kinetic analysis of binding between two proteins [3]. However, fundamental properties of the SPR in flat metal films impose several limitations in SPR biosensors relative to other optical sensing approaches. (1) The decay lengths of propagating surface plasmon polaritons (SPPs) into the dielectric, typically hundreds of nanometers at visible frequencies, are much larger than typical sizes of biomolecules, which occupy only a small fraction of the evanescent field, leading to a less than optimal detection resolution [4]. (2) The large propagation distances (10~45$\mu m$) of SPPs at the smooth metal/dielectric interface lead to crosstalk between adjacent sensing areas, which may limit the achievable spatial resolution (footprint) of SPR biosensors, thus the high-throughput capabilities of multiplexed detections. (3) The complexity and cost of the prism-coupled SPR excitation scheme limit this powerful tool primarily to research laboratories, and are far from ideal for point-of-care devices that can provide instant detection in any place and at any time for personal healthcare. Therefore, the increasing need for sensitive, high-throughput, cost-effective and portable biosensors have driven intense research efforts over the past decade to develop next-generation optical biosensors for clinical and biomedical applications.

Nanoplasmonic biosensors that employ metallic nanoparticles, nanohole or nanoslit arrays to directly couple incident light into SPPs have been proposed to overcome above limitations [4-13]. The localized surface plasmon resonance (LSPR) in metallic nanoparticles exhibits a decay length of tens of nanometers and is highly sensitive to changes close to the surface, making this an attractive approach for biomolecular sensing [4-6]. However, the strong radiative damping and dissipation losses of LSPRs imply broad resonance linewidths, negatively affecting the device sensing performance [4,6]. Nanohole or nanoslit arrays patterned on optically-thick metal films present additional sensing modalities due to the co-existence of propagating SPPs (along metal/dielectric interfaces), LSPRs (excited in individual nanoholes) and plasmonic cavity modes, giving rise to promising phenomena such as extraordinary optical transmission and plasmonic Fano resonances [9-15]. Due to their narrow linewidths (~4*nm*), Fano resonances in metallic nanohole arrays were reported to have high figures-of-merits, surpassing those of commercial SPR sensors [9]. However, results reported to date suffer from the relatively weak resonance intensity and low spectral contrast, which limit the signal-to-noise ratio and thus detection resolution [9-11]. Additionally, several groups have recently demonstrated plasmonic interferometric biosensors utilizing interferences between free-space light and propagating SPPs, and demonstrated compact biosensors with a spectral sensing performance comparable to commercial SPR sensors, as well as a record high intensity-modulation sensitivity [16-18]. Despite the success of this approach, it is limited by low optical transmission through a single nanoslit or nanohole in each sensor unit, affecting the degree of temporal resolution that can be achieved [16-18].

In this work, we present a fundamentally different nanoplasmonic sensor platform exploiting the Fano resonances in one-dimensional (1D) nanogratings patterned on 30*nm*-thick ultrathin Ag films [19-21]. Compared with nanohole or nanoslit arrays patterned on optically-thick metal films [7-13], ultrathin 1D Ag nanogratings exhibit significantly higher optical transmission [19-21], potentially resulting in a strong Fano resonance intensity and high spectral contrast, making them especially suitable for rapid and highly-sensitive sensing applications. Ultrathin nanoplasmonic sensors integrated with PDMS microfluidic channels were fabricated, and their refractive index sensitivity and detection resolution were determined experimentally. The electric field distributions associated with LSPR and SPP modes were determined by Finite-difference time-domain (FDTD) simulations to elucidate the sensor performance and spectral response. The detection, temporal and spatial resolutions determined for the ultrathin plasmonic nanograting sensor are compared to those for previously reported nanoplasmonic biosensors [8-11,16-18]. Our results demonstrate significantly improved temporal and spatial resolution (sensor footprint), and a detection resolution comparable to state-of-the-art nanoplasmonic sensors, creating a promising sensing platform for characterizing the real-time kinetics of surface binding events with a high multiplexing capacity [4].

**Results**

A photograph of an ultrathin (30*nm*-thick) Ag film integrated with a single-channel PDMS microfluidic channel (made via soft lithography) is shown in Fig.1 (a). E-beam evaporation (Indel system) was used to deposit a 2*nm*-thick titanium film and subsequently a 30*nm*-thick Ag film onto a pre-cleaned standard glass slide (Fisher Scientific). Due to the transparency of ultrathin Ag film in the visible regime [21], the background patterns can be clearly seen through the semi-transparent device. In the central area of Ag film (black box), focused ion beam (FIB, FEI Dual-Beam System 235) milling was used to fabricate 1D nanogratings with a period of 420*nm* and slit-width 110*nm*, as shown in Fig. 1(b). After the FIB milling, the sensor chip was cleaned using oxygen plasma (PX-250, March Instruments) and clamped to the PDMS microfluidic channel. Optical measurements were performed in a simple *collinear* transmission geometry using an Olympus IX81inverted microscope [17,21]. A white light beam from a 100*W* halogen lamp illuminated the sample surface through a 50× objective lens, with a spot size of ~50×50$\mu m^2$. The field and aperture diaphragms of the microscope were both closed to obtain a nearly collimated light beam. A linear polarizer was used to limit the incident light to transverse magnetic (TM) polarization. The transmitted light was collected through a 40× objective lens (numerical aperture, NA=0.6), and then coupled to a portable spectrometer (Ocean Optics, USB4000). A CCD camera (COOKE SensiCam QE) was employed to record the position of nanogratings and the illuminated spot.

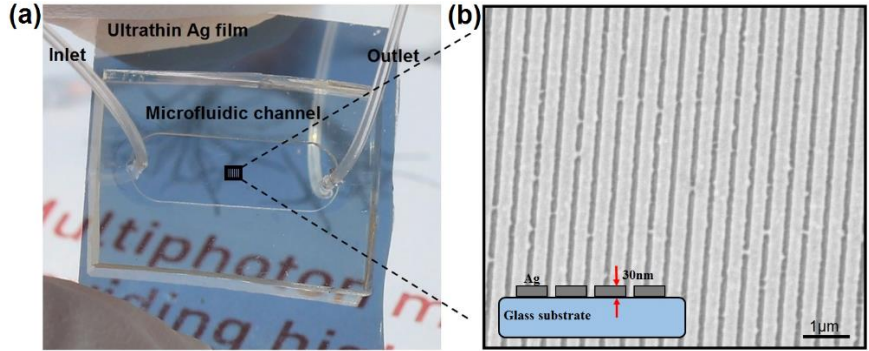

Fig.1. (a) A photograph of an ultrathin (30*nm*-thick) Ag film integrated with a PDMS microfluidic channel. The assembled device is semitransparent, showing the background patterns. 1D nanogratings are fabricated in the center area of Ag film (black box). (b) A scanning electron microscope (SEM) image of 1D nanogratings patterned on the ultrathin Ag film, with a period of 420*nm* and slit-width 110*nm*. Scale bar, 1*μm*. Inset, a cross-section illustration of the ultrathin Ag nanogratings on a glass substrate.

Fig.2 (a) and (b) show the measured and simulated TM polarized optical transmission spectra for light normally incident on a 30*nm*-thick Ag nanograting in a water environment. The period and slit widths of the nanogratings are 420*nm* and 110*nm*, respectively. The spectra were normalized to the reference spectrum of an open aperture with the same area as the nanograting. The transmission spectrum contains two Fano-type asymmetric profiles, each containing a maximum and minimum that are attributed to the constructive and destructive interferences between the broadband LSPR in the nanoslits and two narrowband propagating SPPs on the top (Ag/water) or bottom (Ag/ glass) metal/dielectric interfaces [10, 11, 22-25], respectively. For 1D nanogratings, the momentum matching condition between incident photons and SPPs is fulfilled when the Bragg coupling condition $k_{spp} = 2\pi/\lambda_0 \times \sin\theta + i\,2\pi/P$ is met [26]. Here $k_{spp}$ is the wave-vector of SPPs, $\lambda_0$ incident wavelength, $\theta$ incident angle, $i$ is the grating order, $P$ is the period of nanogratings. The wave-vector relations for SPPs at the metal/dielectric interface is $k_{spp} = 2\pi/\lambda_0 \times \sqrt{\varepsilon_m \varepsilon_d / (\varepsilon_m + \varepsilon_d)}$ ($\varepsilon_m$ and $\varepsilon_d$ are the dielectric constants of metal and dielectric material) [26]. For normally incident light $\theta = 0^0$, the theoretical resonance wavelengths of SPPs are $\lambda_1$=596*nm* and $\lambda_2$=665*nm* at Ag/water and Ag/glass interfaces, respectively, which agree well with the measured and simulated transmission minima [27], as shown in Fig.2 (a) and (b). The spectra differ significantly from the Fano resonances observed in nanoslit or nanohole arrays patterned on optically-thick metal films, which exhibit an extremely low resonance intensity (~1% peak transmission) and modulation ratio [9-11]. The transmission peak in the Fano-type spectrum of the ultrathin Ag nanogratings studied here increased to nearly 40%. The increase in transmitted intensity and modulation ratio at resonance can be attributed to the strong SPPs excited by light transmitted through the nanoslits in the ultrathin metal film. The resonant amplitude of the SPPs is comparable to that of broadband LSPRs, leading to a strong interference between these two modes. This interference is expected to degrade with increasing film thickness, as the reduced transmission through nanoslits (or nanoholes) results in a much lower SPP mode intensity and Fano resonances with a low resonance intensity and modulation ratio [25]. The proposed ultrathin nanostructures with strong Fano resonances create an opportunity to achieve sensing with high signal-to-noise ratio and improved detection resolution.

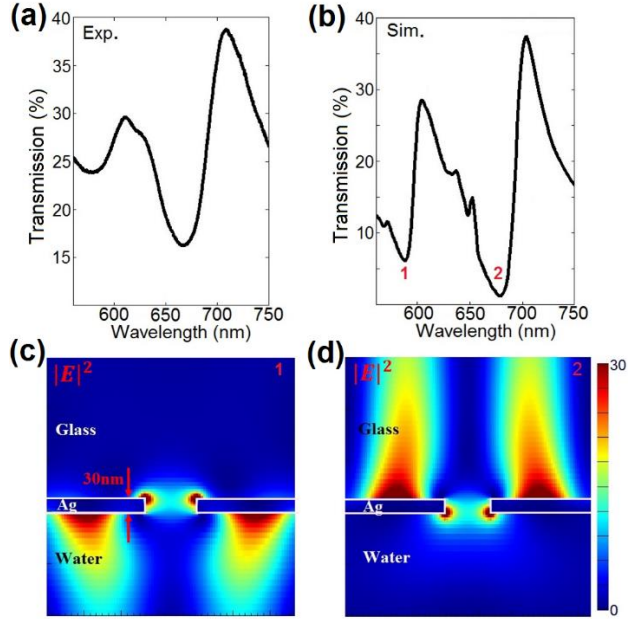

Fig.2. (a) Measured and (b) simulated optical transmission spectra (TM polarization) through 30*nm*-thick Ag nanogratings with a period of 420*nm* and slit-width 110*nm*. Two transmission minima are indicated by the numbers 1 and 2 in (b). (c-d) Electric field $|E|^2$ distributions at two different resonance wavelengths, corresponding to transmission minima 1 and 2 in (b), respectively.

In order to provide a clearer understanding of the physical mechanisms, Fig.2 (c) and (d) plot total electric field $|E|^2$ distributions at two resonance wavelengths corresponding to the transmission minima 1 and 2 in Fig.2 (b), respectively. Fig.2 (c) clearly shows LSPR modes (with a decay length of tens of nanometers) around the corners of the nanogratings at the glass/Ag interface, and SPP modes (with a decay length of hundreds of nanometers) at the water/Ag interface. This pattern is reversed in Fig.2 (d), where the field distributions of LSPR and SPP modes are located at the water/Ag and glass/Ag interfaces, respectively. The electric field distributions clearly demonstrate the co-existence of both LSPR and SPP modes at different resonance wavelengths. The LSPR and SPP modes interact with each other and generate strong Fano resonances with asymmetric spectral profiles [22-24]. The fields of the LSPR mode, which are concentrated within tens of nanometers at the metal surfaces, have the potential to probe surface binding events more sensitively than the propagating SPP modes, which penetrate hundreds of nanometers into the surrounding dielectric and may be more effective in monitoring bulk solution refractive index change. By properly selecting the wavelength, it is possible to excite either LSPR or SPP modes at the water/Ag interface, as shown in Fig.2 (c) and (d), and help to differentiate surface and bulk refractive index changes through their different decay lengths. This unique feature would significantly improve the ability of sensors to detect target biomolecules in complex solutions, such as blood samples [28].

To calibrate the sensor sensitivity and detection resolution, we integrated the fabricated sensor chip with a PDMS microfluidic channel and injected a series of glycerol–water solutions (0, 3, 6, 9, 12, and 15% glycerol concentration) with measured refractive index (RI) ranging from 1.331 to 1.352 (J. A. Woollam, V-VASE). As shown in Fig.3 (a), the non-normalized transmission spectrum red-shifts monotonically with increase in the RI of liquid solutions. The multiple maxima and minima in the non-normalized transmission spectrum arise from interferences of LSPR and SPPs (Fano resonances), as discussed above. The resonance wavelengths of LSPR and SPPs depend on the grating period, film thickness, slit-width, as well as the RI of substrate and surrounding medium. Thus, changes in the local RI of surrounding medium (e.g. glycerol–water solutions) shifts the positions of these spectral features, and can be monitored in real-time by continually recording these spectral features.

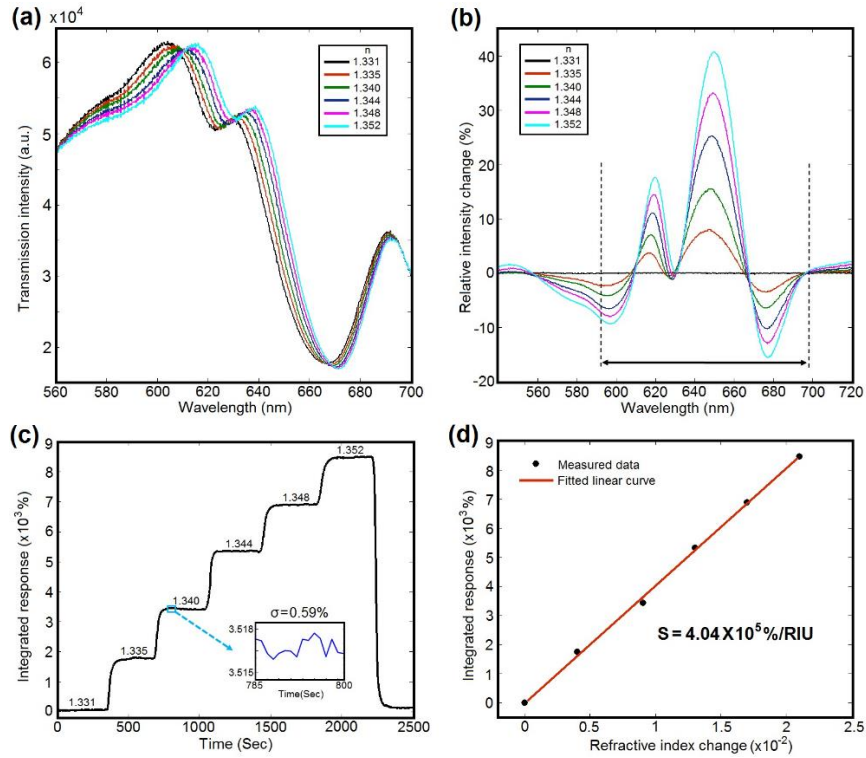

Fig.3. Refractive index sensing using the proposed nanoplasmonic sensors consisting of ultrathin Ag nanogratings (period 420*nm* and slit-width 110*nm*). (a) The measured optical transmission spectra of the sensor in water and 3, 6, 9, 12, 15% glycerol–water mixtures with RI ranging from 1.331 to 1.352, as shown in the inset box. (b) The relative intensity changes $(I(\lambda) - I_0(\lambda))/I_0(\lambda) \times 100\%$ for liquids with different RI. The black dashed lines indicate the integration region. (c) The integrated sensor response as a function of time. The inset shows the noise level of the sensor response over 15 data points with the time interval of 1 *sec*. (d) Extracted sensor output for liquids with different RI. The red line is the linear fit to the measured data (black dot), showing good sensor linearity.

To exploit this sensing principle, we employ a multispectral data analysis method, which makes use of more SPP-mediated spectral information than the traditional single-peak tracking method, and significantly improves the detection signal-to-noise ratio and sensor resolution [16-18, 29]. This method integrates the magnitude of the relative intensity changes $(I(\lambda) - I_0(\lambda))/I_0(\lambda)$ as a function of wavelength, where $I_0(\lambda)$ and $I(\lambda)$ are the transmitted intensities at wavelength $\lambda$ in water and glycerol–water mixtures, respectively. The integrated response (IR) of the sensor can be expressed as: $\mathrm{IR} = \sum_{\lambda_1}^{\lambda_2}(|I(\lambda) - I_0(\lambda)|/I_0(\lambda)) \times 100\%$, where $\lambda_1$ and $\lambda_2$ define the wavelength range for integration. Fig.3 (b) shows the relative intensity changes in percent for liquids of different refractive indices. The signals are most prominent in the spectral range from 590*nm* to 700*nm*, as indicated by the black dashed lines. The magnitude of relative intensity changes was integrated within this spectral range, providing the sensor output IR as a function of time, as shown in Fig.3 (c). The extracted IR values are plotted in Fig.3 (d) as a function of the refractive index, showing that the change in IR is approximately proportional to the increase in refractive index. The linear fit to the data points yields a sensor sensitivity of $4.04 \times 10^5$%/RIU. The inset of Fig.3 (c) indicates the noise level of the IR signal, which exhibits a standard deviation of 0.59% over 15 data points with a time interval of 1 *sec*. This corresponds to a bulk refractive index resolution of $1.46 \times 10^{-6}$ RIU (*i.e.*, $0.59\%/(4.04 \times 10^5 \%/\text{RIU})$).

Table 1. Comparison of detection, temporal and spatial resolutions of state-of-the-art nanoplasmonic biosensors.

| State-of-the-art Nanoplasmonic sensors | Detection Resolution (RIU) | Temporal Resolution (*sec*) | Spatial Resolution ($\mu m^2$) |
|---|---|---|---|
| **Nanogratings on optically-thin Ag film** | $1.46 \times 10^{-6}$ | 1 | $50 \times 50$ |
| Nanohole arrays on optically-thick Au film [8] | $3.1 \times 10^{-6}$ | 2 | $200 \times 200$ |
| Interferometric nanoring-hole arrays [17] | $0.8 \times 10^{-6}$ | 10 | $150 \times 150$ |
| Nanogratings on optically-thick Au film [10, 11] | $1.74 \times 10^{-5}$ | 30 | $\sim 150 \times 150$ |
| Quasi-3D plasmonic crystals [29] | $1 \times 10^{-5}$ | 90 | $1000 \times 5000$ |

Table 1 compares the measured detection, temporal, and spatial resolutions of the optically-thin Ag nanogratings studied here to those of other state-of-the-art nanoplasmonic sensors. As discussed in Ref. [30], there are inherent tradeoffs between these quantities, which need to be carefully balanced to achieve optimal performance of nanoplasmonic sensors. Note that the observed detection resolution for optically-thin Ag nanogratings ($1.6 \times 10^{-6}$ RIU) is among the lowest reported for state-of-the-art nanoplasmonic sensors. Our results compare very well with that reported for optically-thick nanohole arrays ($3.1 \times 10^{-6}$ RIU) [8], but also have 2× better temporal resolution and an order of magnitude improvement in spatial resolution ($50 \times 50$ $\mu m^2$ *v.s.* $200 \times 200$ $\mu m^2$). The optically-thin nanogratings exhibit a slightly higher detection resolution than ring-hole array sensors ($0.8 \times 10^{-6}$ RIU), but have approximately an order of magnitude improvement in both temporal (1 *sec* *v.s.* 10 *sec*) and spatial resolution ($50 \times 50$ $\mu m^2$ *v.s.* $150 \times 150$ $\mu m^2$). In our real-time measurements, $N_t=200$ spectrum frames are averaged with an integration time of $\tau=5$ *msec* for each spectrum, yielding a temporal resolution of $T = N_t \times \tau = 1$ *sec*. This is much shorter than that of nanoplasmonic sensors using ring-hole arrays (T=10 *sec*) [17], 1D nanogratings patterned on optically-thick metal films (T=30 *sec*) [10, 11], and quasi-3D plasmonic crystals (T=90 *sec*) [29]. This improved temporal resolution is attributed to the high resonant optical transmission (~40%) of the ultra-thin Ag nanogratings, which leads to a large photon flux on each detector pixel, permitting a rapid integration time $\tau=5$ *msec*. The temporal resolution T=1 *sec* is comparable to that of commercial SPR sensors, and is sufficient for measuring the binding kinetics of most antibodies.

The temporal resolution, which is important for time-resolved sensing systems, can be further improved by averaging over a smaller number of spectra (e.g. $N_t=50$ or 100 frames) with only a modest increase in sensor noise ($\sigma \propto 1/\sqrt{N_t}$) and correspondingly, a slightly poorer detection resolution [17,30]. The temporal resolution could also be improved by enlarging the sensor footprint (currently $50 \times 50$ $\mu m^2$) to increase the overall detected photon flux and allow a shorter integration time $\tau$, but this would be at the expense of spatial resolution, which is critical for high-throughput multiplexed sensing. Shorter integration times are desired to reduce the effects of shot noise and dark noise in the detector (dominant noise sources in optical biosensor systems), and hence improve the signal-to-noise ratio and detection resolution [8, 31]. Note that although optically-thick nanohole and ring-hole arrays employed larger sensor footprints ($200 \times 200$ $\mu m^2$ and $150 \times 150$ $\mu m^2$) to increase the overall photon flux, the temporal resolution and integration time $\tau$ were still quit long because of the inherently low transmission of those structures. The interferometric ring-hole arrays utilized a $\tau=50$ *msec* integration time, yielding a temporal resolution of T= $200\times50$ *msec*=10 *sec* [17], while nanohole arrays employed a 10*msec* integration time, giving a temporal resolution of T=$200\times10$ *msec*=2 *sec* [8].

The observed improvements in sensor performance result from the high-contrast, narrow bandwidth, and large amplitude of the Fano-type spectrum profiles, as well as the advanced data analysis method employed. Several potential improvements could further enhance the signal-to-noise ratio and sensor resolution. These include: (1) increasing the size of ultra-thin nanogratings using large-area nanofabrication techniques (e.g., optical interference nanolithography [31-37]); (2) reducing the SPP loss by employing ultra-smooth metal films with larger grain sizes (e.g., using template stripping methods) [8,38,39]; (3) using a detector with a higher saturation level and faster frame rates to reduce the shot noise; (4) integrating temperature controllers with both the sensor chip and light source to decrease the noise caused by temperature fluctuations [28]; and (5) optimization of microfluidic channels and analyte mass transport to the sensing surface [4,40], etc.

## Conclusion

In summary, we have demonstrated a nanoplasmonic sensor platform employing Fano resonances in 1D nanogratings patterned on ultrathin Ag films. We observed Fano-type asymmetric spectral profiles with large transmission amplitude (~40%), high modulation ratio, and narrow line-widths. In addition to the simple collinear sensor design, which employs a standard microscope, a portable spectrometer, and PDMS microfluidic channels, this sensor system achieved a detection resolution of $1.46\times10^{-6}$ RIU with a high temporal resolution of 1 *sec*, and a miniaturized sensor footprint of $50\times50$ *$\mu m^2$*. Not only is the sensor resolution among the lowest reported for state-of-the-art nanoplasmonic sensors, but an order of magnitude improvement is also achieved for both temporal (1 *sec v.s.* 10 *sec*) and spatial ($50\times50$ *$\mu m^2$ v.s.* $200\times200$ *$\mu m^2$*) resolutions. Several potential approaches for further improvement were also discussed. The observed superior temporal, spatial and detection resolutions as well as the simple optical geometry suggests this sensing platform has considerable promise for sensitive, cost-effective, and portable biosensors with a high multiplexing capacity; such sensors would significantly impact point-of-care diagnostics for personal healthcare.

## Acknowledgements

We acknowledge the financial support from National Science Foundation (CBET-1014957).

## Supplementary Materials

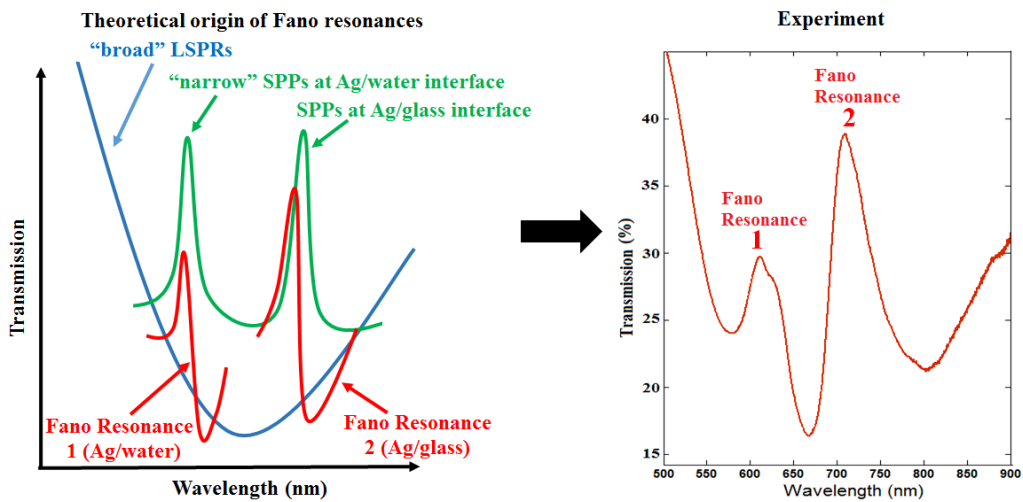

Fig. s1. A theoretical illustration demonstrates Fano-like resonances in the proposed ultrathin Ag nanogratings. The interference of the broadband LSPR resonance and narrowband SPP modes results in the asymmetric Fano-like resonances. The two interfaces (Ag/water and Ag/glass) give rise to two Fano resonance features. The experimental transmission spectrum with two asymmetric Fano-like profiles agree well with the theoretical illustrations.

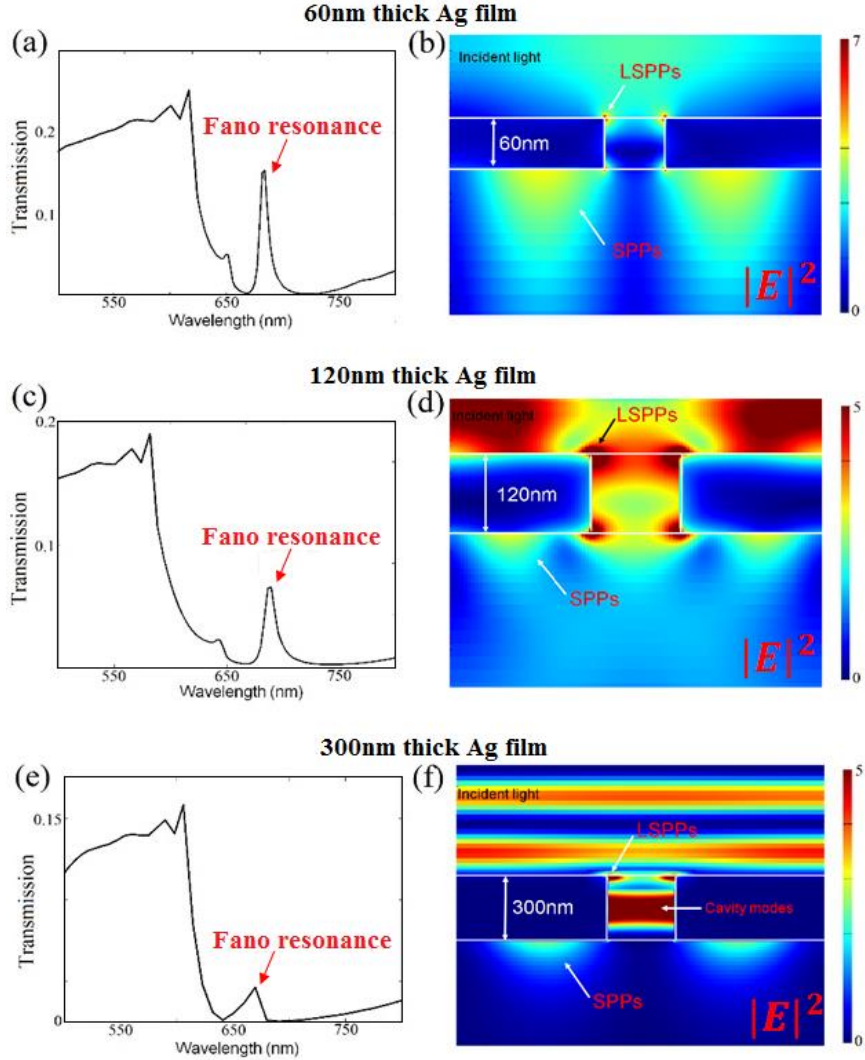

Fig. s2. (a, c, e) Simulated optical transmission spectra (TM polarization) through 60*nm*, 120*nm*, 300*nm*-thick Ag nanogratings with a period of 420*nm* and slit-width 110*nm*. (b,d,f) Electric field $|E|^2$ distributions, corresponding to Fano resonances in (a,c,e), respectively. LSPPs show up on both surfaces as the film becomes thicker from (b) 60*nm* to (d) 120*nm*. And then the cavity mode (f) appears as the film thickness increases up to 300*nm*. As the film thickness increases from 30*nm* to 300*nm*, the intensity of SPPs excited by the light transmitted through the nanoslit becomes weaker, leading to less-effective interference with broadband LSPPs and smaller modulation ratio of Fano resonances.